\documentclass[referee]{aa}
\usepackage[varg]{txfonts}
\usepackage[utf8]{inputenc}
\usepackage[english]{babel}
\usepackage{indentfirst}
\usepackage{graphicx}
\usepackage[]{natbib}
\bibpunct{(}{)}{;}{a}{}{,} 
\usepackage{lineno}

\newcommand{\rem}[1]{}

\begin{document}

\title{Possible structure in the GRB sky distribution at redshift two}
 
\author{Istv\'an Horv\'ath\inst{1}
  \and Jon Hakkila\inst{2}
  \and Zsolt Bagoly\inst{1, 3}  
  }

\offprints{I. Horv\'ath, \email{horvath.istvan@uni-nke.hu}}

\institute{National University of Public Service, Budapest, Hungary,
  \and College of Charleston, Charleston, SC, USA,
  \and E\"otv\"os  University, Budapest, Hungary
  }

\date{Received 12 November, 2013 / Accepted 24 December, 2013}

\abstract
{
Research over the past three decades has revolutionized cosmology 
while supporting the standard cosmological model. However, the cosmological principle of Universal homogeneity and isotropy has always been in question, 
since structures as large as the survey size have always been found each time 
the survey size has increased. Until 2013, the largest known structure in our 
Universe was the Sloan Great Wall, which is more than 400 Mpc long located 
approximately one billion light years away.
}
{Gamma-ray bursts are the most energetic explosions in the Universe. 
As they are associated with the stellar endpoints of massive stars and are 
found in and near distant galaxies, they are viable indicators 
of the dense part of the Universe containing normal matter. 
The spatial distribution of gamma-ray bursts can thus help expose 
the large scale structure of the Universe. } 
{ 
As of July 2012, 283 GRB redshifts have been measured. 
Subdividing this sample into nine radial parts, each containing 31 GRBs,
indicates that the GRB sample having $1.6<z<2.1$ differs significantly from the others in 
that 14 of the 31 GRBs are concentrated in roughly 1/8 of the sky. 
A two-dimensional Kolmogorov-Smirnov test, a nearest-neighbour test, and a 
Bootstrap Point-Radius Method explore the significance of this clustering. 
}
{All tests used indicate that there is a statistically significant clustering of 
the GRB sample at $1.6 < z < 2.1$.  Furthermore, this angular excess cannot be entirely 
attributed to known selection biases, making its existence due to chance unlikely.
} 
{This huge structure lies ten times farther away than the Sloan Great Wall, 
at a distance of approximately ten billion light years. 
The size of the structure defined by these GRBs is about 2000-3000 Mpc, 
or more than six times the size of the largest known object 
in the Universe, the Sloan Great Wall.
}

\maketitle 

\section{Introduction}\label{sec:intro}
The cosmological origin of gamma-ray bursts (GRBs) is well established (e.g. \citealt{MG12}). 
Assuming that the Universe exhibits large-scale isotropy, the same isotropy is also expected for GRBs. 
The large-scale angular isotropy of the sky distribution of GRBs has been well studied over the last few decades. 
Most of these studies have demonstrated that the sky distribution of GRBs is isotropic 
\citep{Briggs96,Teg96,bal98,bal99,mesz00,mgc03,vbh08}.

Some GRB subsamples appear to deviate significantly from isotropy, however. 
\cite{bal98} reported that the angular distributions of short and long GRBs are different. 
\cite{Cline99} found that the angular distributions of very short GRBs are anisotropic, and 
\cite{mgc03} reported that the short GRB class in general deviates from angular isotropy. 
\cite{mesz00} and \cite{li01} wrote that the angular distribution of intermediate duration GRBs is not isotropic. 

In this work we examine the angular distribution of GRBs, and 
we combine this information with the burst radial distribution. 
Our goal is to search the GRB distribution for evidence of large-scale Universal structure.
Since all GRB classes (long, short, and intermediate) are tracers of 
galaxies and matter, our sample consists of all GRBs with known redshift.
As of July 2012, the redshifts of 283 GRBs have been 
determined.\footnote{http://lyra.berkeley.edu/grbox/grbox.php} 
This GRB sample occupies a huge volume, which can presumably provide valuable 
information about Universal large-scale structure.  
To learn more about the properties of the Universe, 
we examine the Copernican principle (homogeneity, isotropy) for this sample.

\section{GRB spatial distribution}\label{sec:GRB spatial distribution}
   By studying the angular distribution of GRBs as a function of distance, 
one can determine sample homogeneity as well as isotropy. 
This sample of GRBs can be subdivided by redshift, resulting in distance 
groupings for which angular information can also be obtained. 
Although sample size 283 GRBs limits our ability to set high-angular 
resolution limits, it can be used for lower-resolution studies. 
We subdivide the sample into five, six, seven, eight, and nine 
different radial subgroups that have sufficient size to justify a statistical study. 

Because of the heterogeneous collection of data in this sample,
the sky exposure function is poorly-characterized for these GRBs, 
making it difficult to test whether all bins have been sampled similarly. 
However, if one assumes similar sampling to first order, 
then one can test whether the two distributions are different or not. 
One common test for comparing two distributions is the Kolmogorov-Smirnov (KS) test. 
However, this test is designed to work with one-dimensional data sets; 
it is hard to use it with data having more than one dimension (such as the angular
position data described here), 
since there is no trivial way to rank in higher dimensions.

A very good summary of how to deal with this problem is given by 
\cite{Lopes08}. For two-dimensional data, \cite{Peacock83} 
suggests that one should use all four possible orderings between ordered pairs
to calculate the difference between the two distributions. 
Since the sky distribution of any object is composed of two orthogonal
angular coordinates, we chose to use this method.

Subdividing the sample by $z$ produces GRB groups whose members are at similar distances from 
us; in other words, their photons come from similar Universal ages. 
This is not true if any group originates from a wide range in $z$. 
Therefore, the dispersion in $z$ needs to be small. 
However, our sample only contains 283 GRBs,  
so the best way to minimize $z$ dispersion is 
to subdivide the data into a larger number of radial bins. 
For that reason we subdivide this sample into five, six, seven, eight, and nine parts. 

When the five groups are compared, following the method discussed in the
next section, there is a weak suggestion of anisotropy in one group. 
When the six groups are compared, there is little sign of any differences between 
the sky distributions of the groups. This is also the case for the samples involving
seven and eight groups, 
but this is not the case when considering the sample containing nine groups.
Therefore, we focus this analysis on the nine redshift groups containing 
GRBs at different redshifts. Each redshift group contains 31 GRBs. 
The minimum redshifts $z_{min}$ that delineate the 9 groups are
0.00 (gr9), 0.41 (gr8), 0.72 (gr7), 0.93 (gr6), 1.25 (gr5), 1.60 (gr4),
2.10 (gr3), 2.73 (gr2), and 3.60 (gr1).

   \section{Two-dimensional Kolmogorov-Smirnov tests}
We used the Peacock methodology to compare the largest absolute differences between
two cumulative angular distributions. Our samples are the 
angular separations between the members in each quadrant in each redshift 
group.\footnote{Typical angular uncertainty for the bursts with observed z less than an
arcsecond, very few have 2-3''. With our data size (283 GRBs) the typical
angular separation for close GRBs is much bigger then a few degrees, so the angular
position errors are negligible in our analysis.} 
When comparing two groups with 31 members 
there are 62x62=3844 division points, and so there are 4x3844 
numbers in each group. For these 15376 pairs, one has to find the largest of their differences.
Comparisons of the nine groups to each other using the 2D K-S test are shown in Table 1. 
Comparing the two farthest groups (gr1, gr2) the largest numerical difference is 9. 
Comparing the two nearest groups (gr8, gr9), the largest numerical difference is 11. 
For the moment, we do not know the precise significance of these numbers; however, 
we can compare them with one another. Table 1 contains the largest number 
in the quadrants for each comparison. Larger numbers indicate larger differences 
between the two groups being compared. Of the six largest numbers, five belong to group4. 
Out of the eight largest numbers, six belong to group4. In other words, six of 
the eight numbers (out of 36) measuring the largest differences between group pairs belong to group4.
	
One can calculate approximate probabilities for the different numbers using simulations. 
We ran 40 thousand simulations where 31 random points are compared with 31 other random points. 
The result contains the number 18 twenty-eight times and numbers larger than 18 ten times, 
so the probability of having numbers larger than 17 is 0.095\%. 
The probability of having numbers larger than 16 is p= 0.0029, of having numbers larger 
than 15 is p= 0.0094, and of having numbers larger than 14 is p=0.0246.  
For a random distribution, this means that numbers larger than 14 correspond to 2$\sigma $ deviations 
and numbers larger than 16 correspond to 3$\sigma$ deviations. 
The probability of having numbers larger than 13 is p=0.057, or 5.7\%, 
which we do not find to be statistically significant.  

\begin{table}[t]\begin{center}
    \hfill{}
    \caption{Results of the 2D KS tests comparing the GRB groups. Boldface means there 
    are significant (more than 2$\sigma $) differences between the sky distributions of the two groups.}
	\begin{tabular}{|l||c|c|c|c|c|c|c|c|c|}\hline 
		 & $z_{min}$ &  gr2  & gr3 & gr4 & gr5 & gr6 & gr7 & gr8 & gr9 \\ \hline \hline 
gr1 & 3.60 &  9  & 9 & {\bf 15} & 11 & 13 & 9 &12 & 8 \\ \hline
gr2 & 2.73 &    & 10  & {\bf 18}  & 7  & {\bf 15}  & 11  & 9  & 12  \\ \hline
gr3 & 2.10 &    &  & {\bf 14} &  9  & 11  & {\bf 14}  & 9  & 10  \\ \hline
gr4 & 1.60 &    &  &  & {\bf 15}  & 10  & {\bf 15}  & {\bf 17}  & 11  \\ \hline
gr5 & 1.25 &    &  &  &  & 13  & 13  & 8  & 10   \\ \hline
gr6 & 0.93 &    &  &  &  &  & 10  & 13  & 8  \\ \hline
gr7 & 0.72 &    &  &  &  &  &  & 10  & 10  \\ \hline
gr8 & 0.41 &    &  &  &  &  &  &  & 11  \\ \hline
	\end{tabular}
	\hfill{}
	\label{tab:err}
\end{center}
\end{table}

As can be seen in Table 1, we have two 3$\sigma$ angular anisotropy signatures. 
In both cases group4 is involved. We also have eight 2$\sigma $ signatures. 
In six cases group4 is involved. However, we do not have a 3$\sigma$ signature 
since we had 36 different pairs to compare. Among 36 different tests one expects 
1.64 2$\sigma $ signatures and no (expected number is 0.09) 3$\sigma$ signatures. Except for cases 
involving group4 we find numbers similar to these random distributions 
(two 2$\sigma $ signatures and no 3$\sigma$ signatures). However the 36 comparisons 
are not independent since we have only nine groups to compare.

\section{Nearest-neighbour statistics}
One can also look for anisotropies using nearest-neighbour statistics. 
Assuming again that the sky exposure is independent of z, one can compare 
the distributions with one another. Since we are not focusing on close pair 
correlations we should calculate not only the nearest-neighbour distances, 
but also the second, third, etc., nearest neighbour distances. 
For all nine groups we calculated the kth (k=1, 2, ... , 30) nearest 
neighbour distance distributions. Since these are one-dimensional 
distributions, a simple Kolmogorov-Smirnov test can be used to determine
significances. 
For eight groups we do not find significant deviation, 
but a KS test for group4 shows significant deviations from 
angular isotropy starting with the sixth-nearest neighbour pairs 
(Fig. 1, top, shows the 10th-neighbour distributions). 

\begin{figure}[h!]\begin{center}
  \resizebox{.95\hsize}{!}{\includegraphics[angle=270]{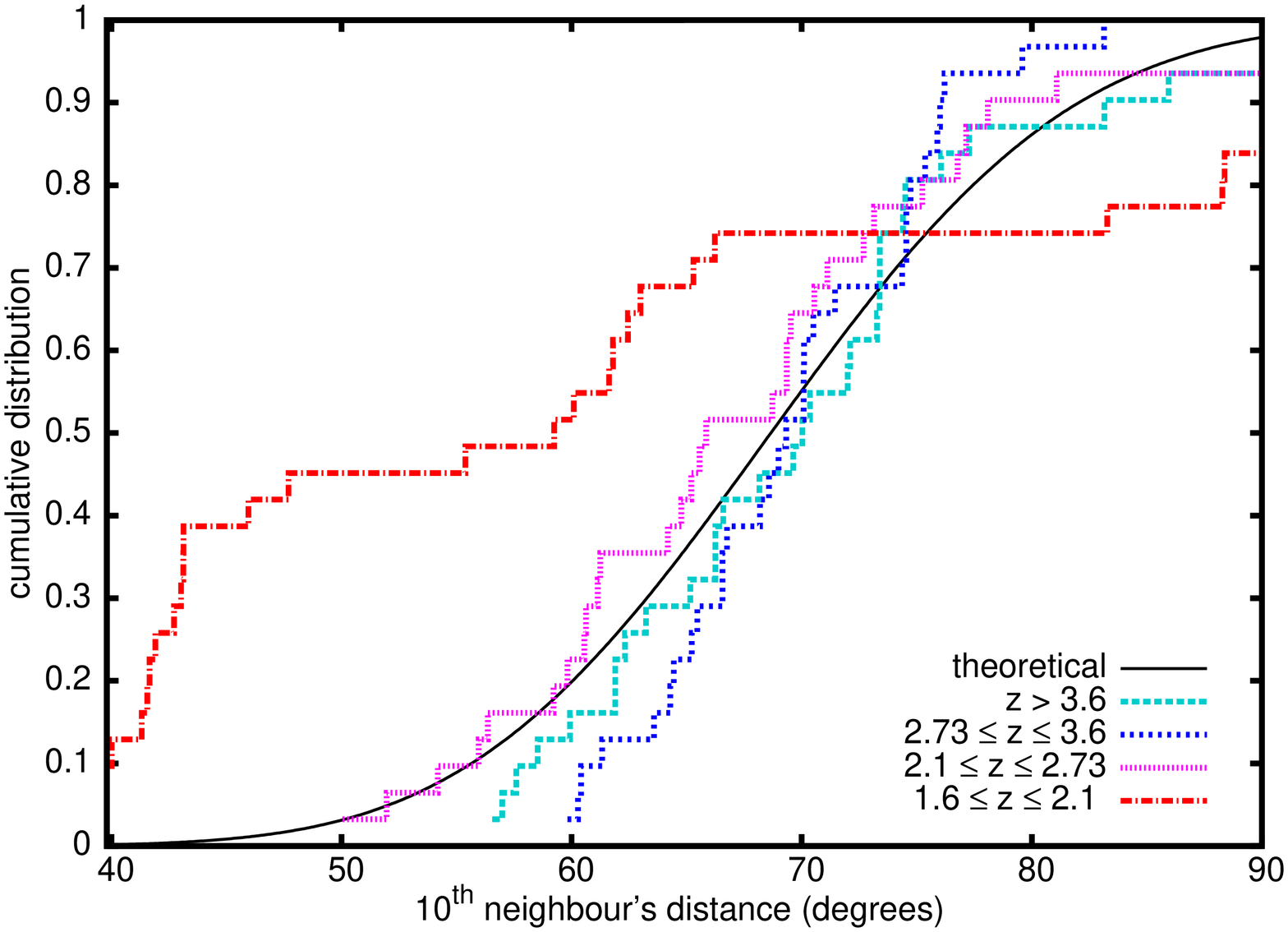}}
  \resizebox{.95\hsize}{!}{\includegraphics[angle=270]{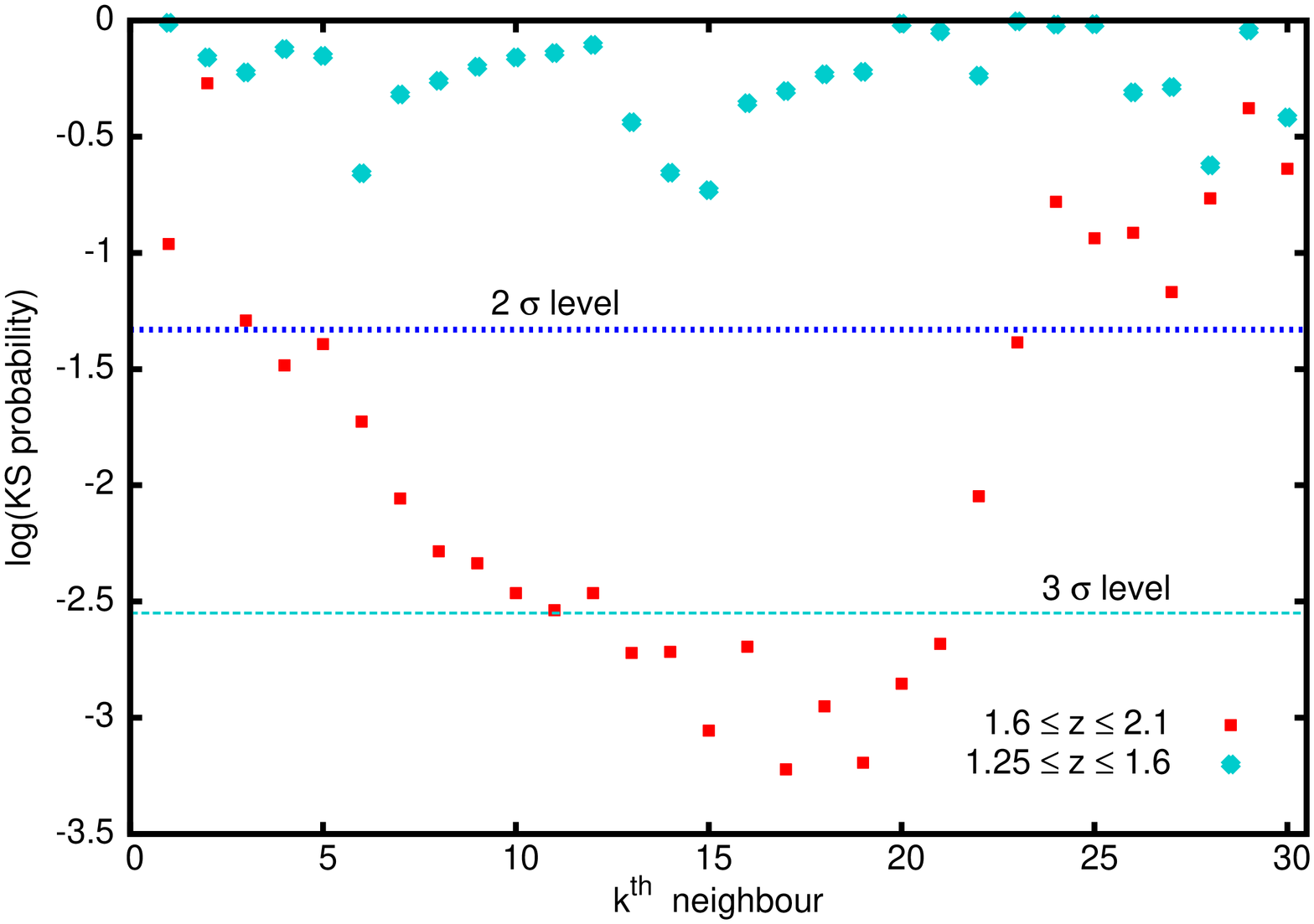}  }
  \caption{\small{
		\textsl{Top:} The 10th-neighbour distribution for the first four groups, red is group4.
				\textsl{Bottom:} K-S neighbour probabilities for group4 (red) and group5 (blue). 
  }}
  \label{fig:090618353}
\end{center}
\end{figure}

The KS probabilities also can be seen in Figure 1. For comparison we also plot the 
group5 probabilities. One can see that 21 consecutive probabilities in group4 
reach the 2$\sigma $  limit and 9 consecutive comparisons reach the 3$\sigma $ limit. 
Of course this does not mean a 27$\sigma $ limit, because the comparisons 
are not independent. One can calculate bimodal probabilities. 
For example 14 out of the 31 GRBs in this redshift 
band are concentrated in approximately 1/8 of the sky (Figure 3). 
The binomial probability of finding this a deviation is p=0.0000055.

\section{Bootstrap point-radius method}
Using a bootstrap point-radius method, we calculated significances 
that the GRB distribution with $1.6 \leq z \leq 2.1$ 
is anisotropic
assuming that the sky 
exposure is essentially independent of $z$.
We chose 31 GRBs from the observed
data set and compared the sky distribution of
this subsample with the sky distribution
of 31 GRBs having $1.6 \leq z \leq 2.1$

To study the selected bursts in two dimensions,
we selected a random location on the
celestial sphere and find how many of the
31 points lie within a circle of predefined angular 
radius, for example, within $10\,^{\circ}$. 
We built statistics for this test
by repeating the process a large number of times (ten
thousand).
From the ten thousand Monte Carlo runs we selected 
the largest number of bursts found within the 
angular circle.

This analysis can be performed with the suspicious 
31 GRB positions and also with 31 randomly chosen 
GRB locations from the observed data. 
There are some angular radii for which
the maximum with the 31 GRBs with $1.6 \leq z \leq 2.1$ is
quite significant. We repeated the process with 31 different randomly
chosen burst positions, and we repeated the experiment 4000 times 
in order to understand the statistical variations of this subsample.
We also performed the same measurement using angular circles
of different radii. 
The frequencies are shown in figure 2.

\begin{figure}[h!]\begin{center}
 
 \resizebox{\hsize}{!}{\includegraphics[angle=270]{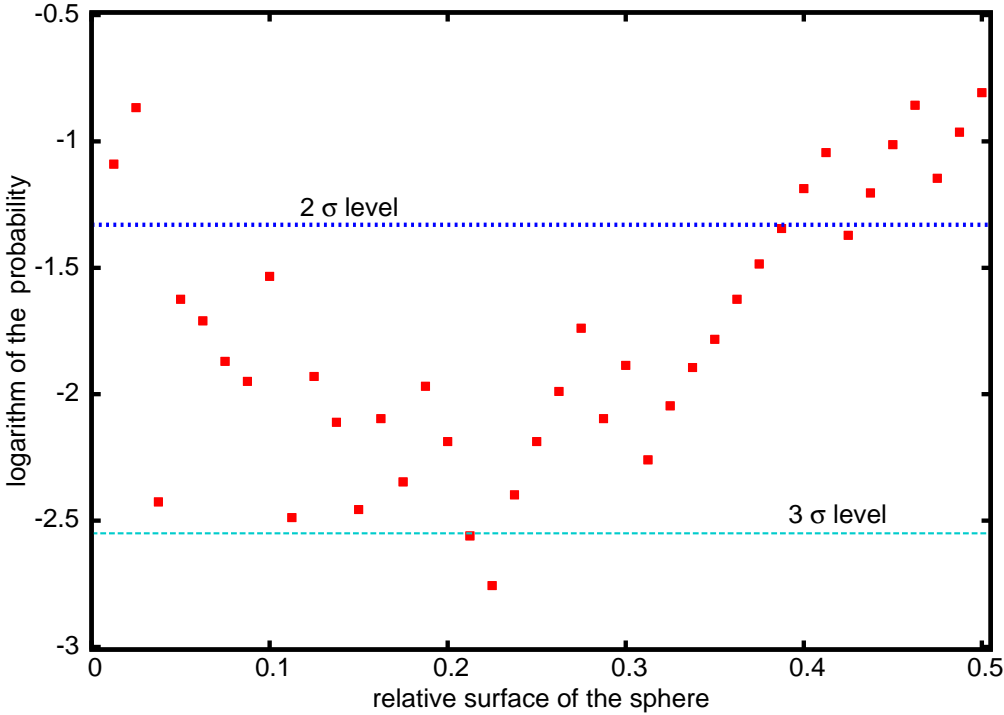} }
 
 \caption{\small{The results of the Monte-Carlo bootstrap point-radius method. 
The horizontal coordinate is the area of the circle in the sky
relative to the whole sky ($4\pi $). The vertical
coordinate is the logarithm of the frequency 
found from the 4000 runs. The two lines show the $2\sigma $ 
and the $3\sigma $ deviations.}}
  \label{fig:0906fd}
\end{center}
\end{figure}

Figure 2 clearly shows that the 15-25 \% of the
sky identified for $1.6 \leq z \leq 2.1$ contains significantly
more GRBs than similar circles at other GRB redshifts.
When the area is chosen to be 0.1125 $\times 4\pi$, 14 GRBs out of the
31 lie inside the circle. 
When the area is chosen to be 0.2125 $\times 4\pi$, 19 GRBs out of the
31 lie inside the circle. 
When the area is chosen to be 0.225 $\times 4\pi$, 20 GRBs out of the
31 lie inside the circle. In this last case only 7 out
of the 4000 bootstrap cases had 20 or more GRBs inside the circle. 
This result is, therefore, a statistically significant (p=0.0018) deviation (the binomial
probability for this being random is less than $10^{-6}$).

\section{Sky exposure effects}
It is important to verify that sampling biases have
not contributed to the previously mentioned angular anisotropies. 
There are two principle causes of angular sampling biases: 
1) GRB detectors often favor triggering on GRBs in some angular directions over others
(sky exposure) 
and 2) GRB redshift measurements, made in the visual/infrared, are
inhibited by angularly-dependent Galactic extinction (e.g. \citealt{hak97}). 

\begin{figure*}[h!]
  \sidecaption \includegraphics[angle=90,width=12.7cm]{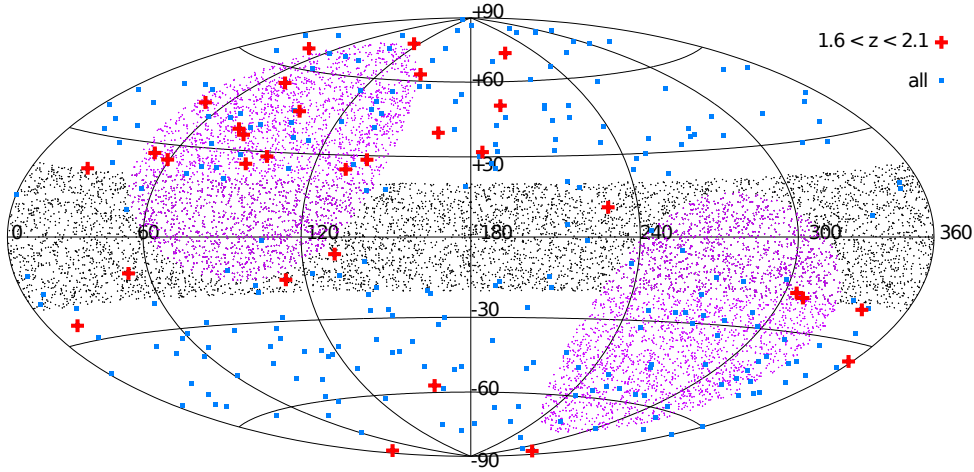} 
  \caption{\small{Sky distribution of 283 GRBs 
  with redshift in Galactic coordinates (blue squares), 
  overlaid by modeled angular biases. The Galactic equator region (lightly 
  mottled) is under-sampled because of Galactic extinction effects on 
  redshift measurements. The ecliptic polar regions (pink mottled) are 
  oversampled relative to ecliptic
  equator regions because of Swift's sky exposure. The anisotropic distribution
  of the 31 GRBs at $1.6 \le z \le 2.1$ (red crosses) is largely unaffected by these 
  biases; sampling biases have not favored their detection.}}
  \label{fig:biases}
\end{figure*}

Anisotropic sampling can be produced when a pointed spacecraft 
observes some sky directions more often than others, or 
when the field of view is blocked (e.g. from Earth occultation, avoidance 
of the Sun to protect instrumentation, or trigger disabling such as 
over the South Atlantic Anomaly). 
Each GRB instrument samples the sky differently, 
which makes the summed sky exposure difficult to identify for our 
heterogeneous GRB sample that has been observed by many instruments since the late 1990s. 
However, 214 of the 283 GRBs in our sample (75.6\%) have been observed 
by Swift, as have 23 of the 31 GRBs in group4 (74.2\%). 
Thus, we assume to first-order that Swift's sky exposure is a reasonable
approximation of the sky exposure of the entire burst sample. 
Swift's sky exposure has recently been published \citep{Bau13};
its primary characteristic is that the regions near the ecliptic poles 
($| \beta | \ge45\,^{\circ}$, where $\beta$ is the ecliptic latitude) have been
observed roughly $1.83$ times more frequently than the region surrounding
the ecliptic equator.

Extinction due to dust from the Milky Way disk causes another angularly-dependent
sampling bias.
Gamma--ray bursts included in our sample have not merely been
detected, but have also had their redshifts measured. Redshift
measurements involve visual (and/or infrared) spectral observations
made during the burst's afterglow phase. Extinction can reduce afterglow 
brightness by several magnitudes or more, making spectral line measurements
more difficult. Close to the Galactic plane, the afterglow may not even be
detected. The details of the process by which redshift measurements are 
affected are difficult to model, as they depend
on the intrinsic luminosity and decay rate of the afterglow, the telescope and 
spectrograph used, the redshift of the burst, the observing conditions at 
the time of detection, and of course the Galactic latitude and longitude of
the burst.

Our first-order model of angular biasing in the sample assumes
that GRB detection rates are enhanced 
by a factor of $1.83$ at high ecliptic latitudes (those with 
$| \beta | \ge45\,^{\circ}$) compared to those near the ecliptic equator.
Second, we account for the spottiness of redshift observations near 
the Galactic plane by assuming that the fraction of low Galactic latitude 
($b < 20\,^{\circ}$) GRBs in group4 is the same as that found
in the other groups (12.9\%), and that this value is a reasonable
representation of the true redshift measurement rate.
These assumptions seems to be reasonable for
the total sample of 283 GRBs. The resulting skymap is shown in Figure 3.

We test the hypothesis that Swift's sky exposure and Galactic
extinction might be responsible for the group4 burst clustering 
by examining the predicted numbers of GRBs in the best-sampled regions 
near the ecliptic poles ($b \ge 20\,^{\circ}$ and  $| \beta | \ge45\,^{\circ}$) 
compared to those in the well-sampled regions near the ecliptic equator
($b \ge 20\,^{\circ}$ and $| \beta | < 45\,^{\circ}$) and to those
in the poorest-sampled region ($b < 20\,^{\circ}$).

The expected numbers of GRBs in each group are 
$5.41$ (best, north), $5.41$ (best, south), $8.09$ (good, north), $8.09$ (good, south), 
and $4.00$ (poor), whereas the actual counts are
$11$ (best, north), $1$ (best, south), $8$ (good, north), $6$ (good, south), 
and $6$ (poor).
This results in a $\chi^2$ probability of $p=0.025$
that this is due to chance. 
This probability indicates that more GRBs
were detected in the well-sampled northern region and fewer were
detected in the well-sampled southern region than were
expected owing to sky exposure and extinction effects.
The $2.5\%$ probability refers only to the fact that 
extinction and exposure effects cannot explain the group4 
anisotropy and do not describe the
strong clustering of GRBs within the
well-sampled northern region.
Finally, we note that the southern depletion is not surprising if the
northern clustering is real: if the cluster is real, then our choice of requiring equal
GRB counts per radial bin allows us to detect group4 bursts in the cluster 
at the cost of detecting them elsewhere.

\section{Summary and conclusion}
Here we report the discovery of a possible large-scale Universal 
structure at a distance of approximately ten billion light years (redshift $z\approx2$). 
The two-dimensional KS test shows a $3\sigma $ deviation.
The bootstrap point-radius method confirms this significance.
The nearest-neighbuor statistics also reach the $3\sigma $ level several
times. Sampling biases cannot explain the
clustering of GRBs at this redshift.

Even though it is widely accepted, 
the cosmological principle of Universal 
homogeneity and isotropy has always been questioned, 
since structures as large as the survey size have been 
consistently found as the survey size has increased. 
In the late 1980s \cite{Gel89} mapped the 
Universe to $z\approx0.03$ and found a 200 Mpc size object, 
which they called the Great Wall. 
In 2005 an object twice this size was reported and was named 
the Sloan Great Wall \citep{Gott05}. Recently,  \cite{clo12} have found a 
large quasar group (LQG) with a length of 1000 Mpc.
In this study we have found a potential structure, 
mapped by GRBs, of about 2000-3000 Mpc size. 
It is interesting to note that the universal star formation, 
and thus also the GRB rate, peaked between redshifts 
$1.6 \le z \le 2.1$ \citep{haye10}.
Since GRBs are a luminous tracer of matter, 
they should be even better tracers at these distances,
where they are more common. 
In other words, if one wants to search for evidence of the 
largest clumps of matter in the universe using GRBs (or 
any luminosity indicator that correlates with the star formation 
rate), then one should search in this redshift range.

This cluster is not apparent in the larger angular distribution of
all detected GRBs. This is not surprising since the cluster
occupies a small radial region coupled with a large angular one.
In the coming years, additional GRB detection
by Swift and new missions such as SVOM, UFFO, and CALET, 
coupled with successful follow-up redshift measurements,
should provide the statistics to confirm or disprove this discovery.

\begin{acknowledgements} 
This research was supported by OTKA grant K77795 and by NASA ADAP grant NNX09AD03G. 
The authors thank the referee for improving the manuscript.
Discussions with L.G. Balázs are also acknowledged.
\end{acknowledgements}

\bibliographystyle{aa} 
\bibliography{horv13}

\hyphenation{Post-Script Sprin-ger}
\begin{thebibliography}{18}
\expandafter\ifx\csname natexlab\endcsname\relax\def\natexlab#1{#1}\fi

\bibitem[{{Bal{\'a}zs} {et~al.}(1998){Bal{\'a}zs}, {M{\'e}sz{\'a}ros}, \&
  {Horv{\'a}th}}]{bal98}
{Bal{\'a}zs}, L.~G., {M{\'e}sz{\'a}ros}, A., \& {Horv{\'a}th}, I. 1998, \aap,
  339, 1

\bibitem[{{Bal{\'a}zs} {et~al.}(1999){Bal{\'a}zs}, {M{\'e}sz{\'a}ros},
  {Horv{\'a}th}, \& {Vavrek}}]{bal99}
{Bal{\'a}zs}, L.~G., {M{\'e}sz{\'a}ros}, A., {Horv{\'a}th}, I., \& {Vavrek}, R.
  1999, \aaps, 138, 417

\bibitem[{{Baumgartner} {et~al.}(2013){Baumgartner}, {Tueller}, {Markwardt},
  {Skinner}, {Barthelmy}, {Mushotzky}, {Evans}, \& {Gehrels}}]{Bau13}
{Baumgartner}, W.~H., {Tueller}, J., {Markwardt}, C.~B., {et~al.} 2013, \apjs,
  207, 19

\bibitem[{{Briggs} {et~al.}(1996){Briggs}, {Paciesas}, {Pendleton}, {Meegan},
  {Fishman}, {Horack}, {Brock}, {Kouveliotou}, {Hartmann}, \&
  {Hakkila}}]{Briggs96}
{Briggs}, M.~S., {Paciesas}, W.~S., {Pendleton}, G.~N., {et~al.} 1996, \apj,
  459, 40

\bibitem[{{Cline} {et~al.}(1999){Cline}, {Matthey}, \& {Otwinowski}}]{Cline99}
{Cline}, D.~B., {Matthey}, C., \& {Otwinowski}, S. 1999, \apj, 527, 827

\bibitem[{{Clowes} {et~al.}(2013){Clowes}, {Harris}, {Raghunathan},
  {Campusano}, {S{\"o}chting}, \& {Graham}}]{clo12}
{Clowes}, R.~G., {Harris}, K.~A., {Raghunathan}, S., {et~al.} 2013, \mnras,
  429, 2910

\bibitem[{{Geller} \& {Huchra}(1989)}]{Gel89}
{Geller}, M.~J. \& {Huchra}, J.~P. 1989, Science, 246, 897

\bibitem[{{Gott} {et~al.}(2005){Gott}, {Juri{\'c}}, {Schlegel}, {Hoyle},
  {Vogeley}, {Tegmark}, {Bahcall}, \& {Brinkmann}}]{Gott05}
{Gott}, III, J.~R., {Juri{\'c}}, M., {Schlegel}, D., {et~al.} 2005, \apj, 624,
  463

\bibitem[{{Hakkila} {et~al.}(1997){Hakkila}, {Myers}, {Stidham}, \&
  {Hartmann}}]{hak97}
{Hakkila}, J., {Myers}, J.~M., {Stidham}, B.~J., \& {Hartmann}, D.~H. 1997,
  \aj, 114, 2043

\bibitem[{{Hayes} {et~al.}(2010){Hayes}, {Schaerer}, \& {{\"O}stlin}}]{haye10}
{Hayes}, M., {Schaerer}, D., \& {{\"O}stlin}, G. 2010, \aap, 509, L5

\bibitem[{{Litvin} {et~al.}(2001){Litvin}, {Matveev}, {Mamedov}, \&
  {Orlov}}]{li01}
{Litvin}, V.~F., {Matveev}, S.~A., {Mamedov}, S.~V., \& {Orlov}, V.~V. 2001,
  Astronomy Letters, 27, 416

\bibitem[{{Lopes} {et~al.}(2008){Lopes}, {Hobson}, \& {Reid}}]{Lopes08}
{Lopes}, R.~H.~C., {Hobson}, P.~R., \& {Reid}, I.~D. 2008, Journal of Physics
  Conference Series, 119, 042019

\bibitem[{{Magliocchetti} {et~al.}(2003){Magliocchetti}, {Ghirlanda}, \&
  {Celotti}}]{mgc03}
{Magliocchetti}, M., {Ghirlanda}, G., \& {Celotti}, A. 2003, \mnras, 343, 255

\bibitem[{{M{\'e}sz{\'a}ros} {et~al.}(2000){M{\'e}sz{\'a}ros}, {Bagoly},
  {Horv{\'a}th}, {Bal{\'a}zs}, \& {Vavrek}}]{mesz00}
{M{\'e}sz{\'a}ros}, A., {Bagoly}, Z., {Horv{\'a}th}, I., {Bal{\'a}zs}, L.~G.,
  \& {Vavrek}, R. 2000, \apj, 539, 98

\bibitem[{{M{\'e}sz{\'a}ros} \& {Gehrels}(2012)}]{MG12}
{M{\'e}sz{\'a}ros}, P. \& {Gehrels}, N. 2012, Research in Astronomy and
  Astrophysics, 12, 1139

\bibitem[{{Peacock}(1983)}]{Peacock83}
{Peacock}, J.~A. 1983, \mnras, 202, 615

\bibitem[{{Tegmark} {et~al.}(1996){Tegmark}, {Hartmann}, {Briggs}, \&
  {Meegan}}]{Teg96}
{Tegmark}, M., {Hartmann}, D.~H., {Briggs}, M.~S., \& {Meegan}, C.~A. 1996,
  \apj, 468, 214

\bibitem[{{Vavrek} {et~al.}(2008){Vavrek}, {Bal{\'a}zs}, {M{\'e}sz{\'a}ros},
  {Horv{\'a}th}, \& {Bagoly}}]{vbh08}
{Vavrek}, R., {Bal{\'a}zs}, L.~G., {M{\'e}sz{\'a}ros}, A., {Horv{\'a}th}, I.,
  \& {Bagoly}, Z. 2008, \mnras, 391, 1741

\end{thebibliography}

\end{document}